\documentclass[aps,prl,twocolumn,showpacs,floatfix,superscriptaddress]{revtex4}
\usepackage{graphicx}

\begin{document}
\flushbottom

\title{Nodeless spin-triplet superconducting gap in Sr$_2$RuO$_{4}$}
\author{H. Suderow}
\affiliation{Laboratorio de Bajas Temperaturas, Departamento de
F\'isica de la Materia Condensada \\ Instituto de Ciencia de
Materiales Nicol\'as Cabrera, Facultad de Ciencias \\ Universidad
Aut\'onoma de Madrid, 28049 Madrid, Spain}
\author{V.  Crespo}
\affiliation{Laboratorio de Bajas Temperaturas, Departamento de
F\'isica de la Materia Condensada \\ Instituto de Ciencia de
Materiales Nicol\'as Cabrera, Facultad de Ciencias \\ Universidad
Aut\'onoma de Madrid, 28049 Madrid, Spain}
\author{I. Guillamon}
\affiliation{Laboratorio de Bajas Temperaturas, Departamento de
F\'isica de la Materia Condensada \\ Instituto de Ciencia de
Materiales Nicol\'as Cabrera, Facultad de Ciencias \\ Universidad
Aut\'onoma de Madrid, 28049 Madrid, Spain}
\author{S. Vieira}
\affiliation{Laboratorio de Bajas Temperaturas, Departamento de
F\'isica de la Materia Condensada \\ Instituto de Ciencia de
Materiales Nicol\'as Cabrera, Facultad de Ciencias \\ Universidad
Aut\'onoma de Madrid, 28049 Madrid, Spain}
\author{F. Servant}
\affiliation{Institut N\'{e}el, CNRS / UJF, 25, Av. des Martyrs,
BP166, 38042 Grenoble Cedex 9, France}
\author{P. Lejay}
\affiliation{Institut N\'{e}el, CNRS / UJF, 25, Av. des Martyrs,
BP166, 38042 Grenoble Cedex 9, France}
\author{J.P. Brison}
\affiliation{CEA, INAC, SPSMS, 38054 Grenoble, France }
\author{J. Flouquet}
\affiliation{CEA, INAC, SPSMS, 38054 Grenoble, France }

\begin{abstract}
We report on tunneling spectroscopy measurements using a Scanning Tunneling Microscope (STM) on the spin triplet superconductor Sr$_2$RuO$_4$. We find a negligible density of states close to the Fermi level and a fully opened gap with a value of $\Delta$=0.28 meV, which disappears at T$_c$ = 1.5 K. $\Delta$ is close to the result expected from weak coupling BCS theory ($\Delta_0$=1.76k$_B$T$_c$ = 0.229 meV). Odd parity superconductivity is associated with a fully isotropic gap without nodes over a significant part of the Fermi surface.\end{abstract}

\pacs{74.20.Rp, 74.70.Pq, 74.25.Jb} \date{\today} \maketitle

Superconductivity in the ruthenate compound Sr$_2$RuO$_4$ has been puzzling many scientists during the past decade\cite{Maeno94}. A number of experiments, reviewed in Ref.\cite{Mackenzie03}, show that this superconductor could become the metallic analog to superfluid $^3$He. Strong indications that this material may be indeed an odd parity superconductor have been repeatedly found since the first report for the absence of a change in the Knight shift in NMR experiments\cite{Ishida98}. Sr$_2$RuO$_4$  crystalizes in the layered perovskite structure common to cuprates \cite{Maeno94}, it has a critical temperature $T_c$ of 1.5 K, and is a good metallic system with a relatively simple Fermi surface consisting of three nearly cylindrical sheets, all of them deriving from the Ru 4d orbitals, and mass renormalization factors between 3 and 5.5\cite{Mackenzie96}. Identifying the structure of the superconducting gap over the Fermi surface is of prime importance to understand spin triplet superconductivity in this material. In high quality samples, the electronic contribution to the specific heat or the thermal conductivity vanishes at the lowest temperatures\cite{Nishizaki99,Izawa01,Tanatar01b,Ishida00,Lupien01}. The curves do not show activated behavior, which has been taken as an evidence for the presence of nodes somewhere on the Fermi surface \cite{Mackenzie03}. A remarkable zero bias anomaly compatible with unconventional superconductivity was observed in point contact spectroscopy experiments. However, the reported gap values ranged up to five times the result expected from weak coupling BCS theory\cite{Laube00}. The first published tunneling spectroscopy experiments using STM have given curves showing very smeared superconducting features, with a high amount of excitations at the Fermi level (85 \% of the normal state density of states), and gap values also well above those expected within weak coupling BCS theory\cite{Upward02}. A lattice distortion, which may produce ferromagnetic behavior close to the surface, was found in room temperature STM experiments\cite{Matzdorf00}. Beautiful atomic resolution images of the SrO plane were obtained in in-situ cleaved crystals at very low temperatures in Ref.\cite{Barker03}. However, no sign of the superconducting gap was reported. So it is clear that, although tunneling and point contact spectroscopy are in principle powerful tools to measure the superconducting gap \cite{Fischer07}, these techniques have been until now unable to bring uncontroverted answers to some of the most important opened issues in Sr$_2$RuO$_4$. Namely, what is the size of the superconducting gap, and which kind of nodes, if any, are found over the Fermi surface ? In previous work, we have shown that the superconducting gap can be measured in small superconducting isolated grains \cite{Rubio01}. The tip can be easily moved in-situ to search for superconducting features in many different grains and positions. Moreover, we have also introduced the use of superconducting tips of Al for STM experiments through a detailed study of NbSe$_2$, which included atomic resolution as well as the observation of the vortex lattice \cite{Guillamon07}. Here we report about tunneling spectroscopy results in single grains of Sr$_2$RuO$_4$ using a superconducting tip of Al, where we have been able to obtain new important information about the gap opening in this material.

\begin{figure}[ht]
\includegraphics[angle=270,width=9cm,clip]{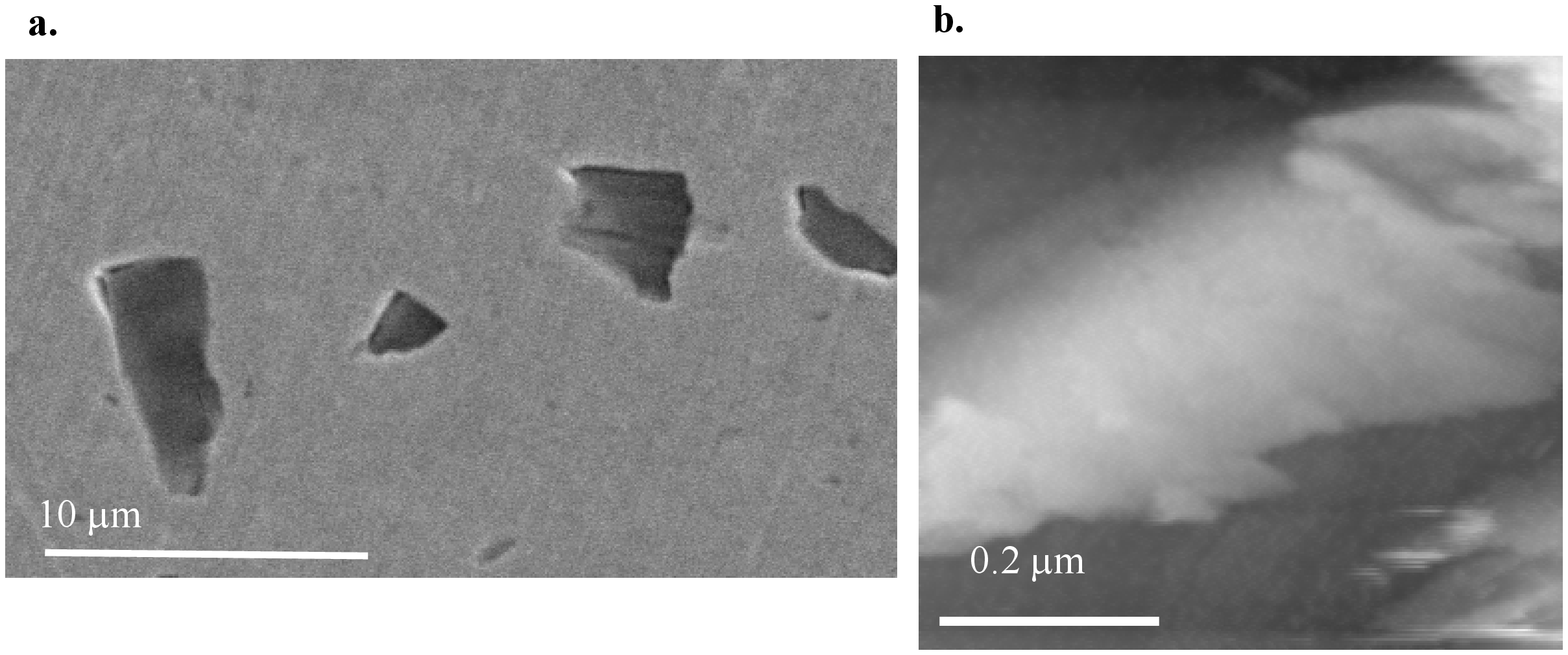}
\vskip -2.8cm \caption{In a. we show a Scanning Electron Microscope image of several grains of Sr$_2$RuO$_4$. In b. we show a single small grain as viewed with the STM.} \label{Fig1}
\end{figure}

We use a STM/S in a dilution refrigerator, tested previously in measurements in different systems, among them pure Al \cite{Rodrigo04,Suderow04,Guillamon07}. The set-up cools down tip and sample to 150 mK and has a resolution in energy of 15 $\mu$eV. The sample was mounted on the sample holder following the procedure described in Ref.\cite{Rubio01}. A small (1 mm $\times$ 1 mm $\times$ 1 mm) Sr$_2$RuO$_4$ single crystal was crushed between two flat synthetic ruby, resulting in many small single crystallites of Sr$_2$RuO$_4$ that were subsequently gently pressed, using another flat ruby, onto a clean surface of high purity Al. Immersing the resulting sample into an ultrasound bath of acetone removed grains that were not firmly attached to the Al surface. We find that most of the surface is free of Sr$_2$RuO$_4$, except for small micrometer size isolated grains with large Al areas in between (Fig.\ref{Fig1}). The Al tip was positioned at room temperature over the Al surface, in between grains. Subsequently, at low temperatures, the tip was cleaned using the repeated indentation procedure described in Refs.\cite{Rodrigo04b,Guillamon07}.  Using a x-y table that allows for precise coarse movement of the tip over the sample, we searched in-situ for isolated Sr$_2$RuO$_4$ grains. Specificities and advantages of the use of superconducting tips have been discussed in Refs.\cite{Stip1,Rodrigo04b,Kohen06}. Note in particular that, with typical tunneling resistances of the order of the M$\Omega$, the Josephson coupling energy easily falls below a few tens of mK, and the Josephson current appears to be strongly suppressed \cite{Stip1,Rodrigo04b,Kohen06}. Therefore, here we only focus on the features of the tunneling experiment related to the superconducting densities of states of tip and sample.

\begin{figure}[ht]
\includegraphics[width=8cm,clip]{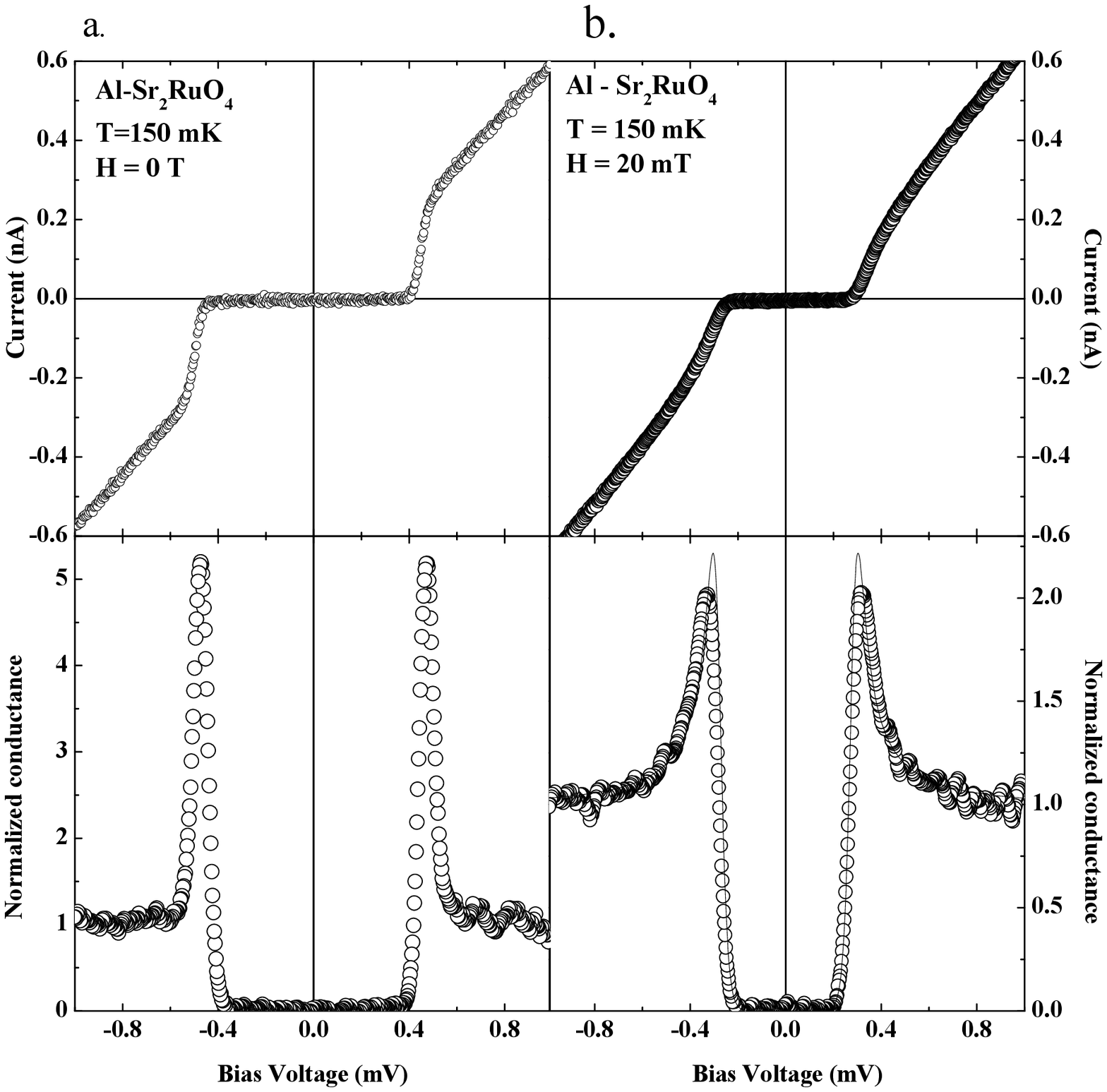}
\vskip -4cm \caption{Tunneling current (top panels) and corresponding tunneling conductance (bottom panels) between a tip of Al and a sample of Sr$_2$RuO$_4$. In b (right panels), a magnetic field of 20 mT has been applied to destroy superconductivity on the tip. Line is a fit to BCS theory, using a superconducting gap of $\Delta_S=\Delta_{Sr_2RuO_4}$ = 0.28 meV and T = 0.15 K.} \label{Fig2}
\end{figure}

Equation \ref{Equ1} gives the well known expression for the tunneling current as a function of the bias voltage $I(V)$ at a given position, valid for voltages below some tens of mV (we assume that the tip and sample wavefunction coupling gives simply an energy independent proportionality factor).

\begin{equation}
I(V) \propto \int dE [f(E-eV)-f(E)] N_T(E-eV) N_S(E)
\label{Equ1}
\end{equation}

Here $f$ is the Fermi function, and $N_T$ and $N_S$ are, respectively, the densities of states of tip and sample. At zero temperature, and using BCS expressions for both tip and sample densities of states, it is easy to find that $I(V)$ is zero for $|V|<\Delta_T+\Delta_S$, and has a very sharp increase for $|V|=\Delta_T+\Delta_S$. This leads to a zero tunneling conductance $\sigma(V)$ below $\Delta_T+\Delta_S$ and a sharp peak at $\Delta_T+\Delta_S$. $I(V)$ grows with temperature at low voltages, with the tail of the Fermi function, and an additional peak appears in $\sigma(V)$ at $V=\pm(\Delta_S-\Delta_T)$.

In our experiment, at 150 mK ($\approx T_c/10$), we find a tunneling conductance which is zero close to the Fermi level, and a very well defined and sharp peak at $\Delta_T+\Delta_S$=0.46 meV (Fig.\ref{Fig2}a). The superconducting gap of pure Al, measured in the same set-up when tunneling onto the free surface of Al, is of $\Delta_T$=0.18 meV. Therefore, the superconducting gap of Sr$_2$RuO$_4$ is of $\Delta_S$=0.28 meV. Note on the other hand that the peak in $\sigma(V)$ is sharp and well defined, yet it has a finite width at half height of about 80 $\mu$eV. Part of it is due to the finite resolution in voltage of our set-up (15 $\mu$eV). However, another significant part should be due to deviations from fully isotropic gap behavior in Sr$_2$RuO$_4$. To further study this point, we applied a small magnetic field to destroy superconductivity in the Al tip. As is well known, superconductivity in Al is destroyed at very weak magnetic fields, unless the tip is specifically shaped to a small cone like structure as shown and discussed in Refs.\cite{Rodrigo04b,Guillamon07}. Here, the tip has not been specifically prepared, and we find that, above about 10 mT, superconductivity in the tip abruptly disappears. When the tip is normal, the tunneling conductance $\sigma$(V) is simply the temperature smeared density of states of the sample. The result of our experiment at 20 mT and 150 mK is shown in Fig.\ref{Fig2}b, together with a fit to BCS density of states using $\Delta_S$=0.28 meV. As is clearly seen, the fit reproduces well the experiment, except for a small discrepancy at the quasiparticle peak. This discrepancy can be easily removed by allowing for a Gaussian distribution of values of the superconducting gap around $\Delta_S$=0.28 meV, with a total width of around 60 $\mu$eV. This gap anisotropy is the remaining contribution to the finite width of the quasiparticle peak observed at zero field when the tip is superconducting (bottom panel of Fig.\ref{Fig2}a). Therefore, the superconducting density of states of Sr$_2$RuO$_4$ very closely follows expectations for a fully isotropic gap, with a distribution of values of the superconducting gap centered at $\Delta_S$=0.28 meV with a width of 60 $\mu$eV.

\begin{figure}[ht]
\includegraphics[width=8cm,clip]{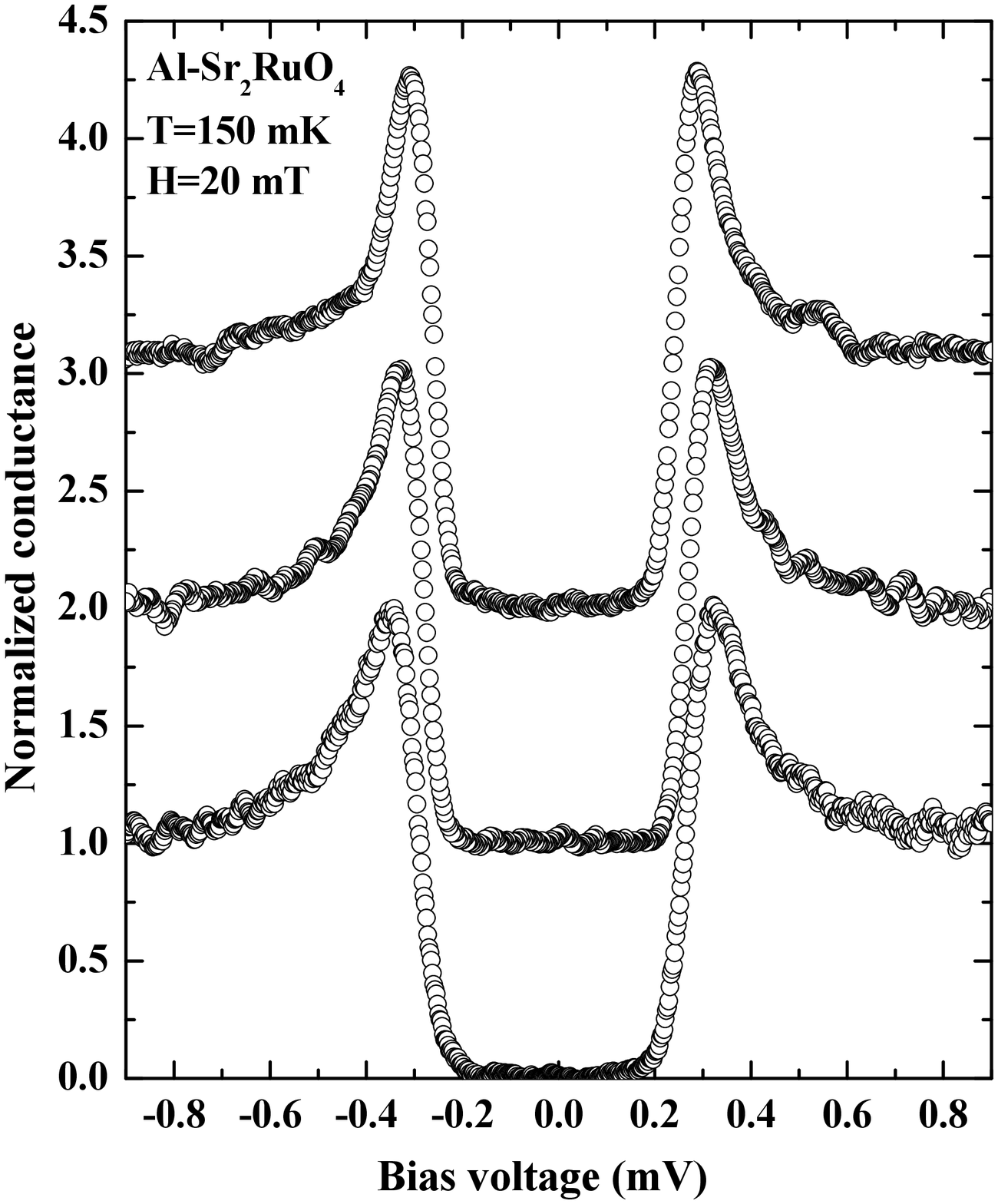}
\vskip -1.5cm \caption{Tunneling conductance curves found at different positions (shifted for clarity).}
\label{Fig3}
\end{figure}

These results have been reproducibly found in different grains, as well as in different positions of a single grain. As an example, we show further tunneling conductance curves obtained with a small magnetic field in different grains in Fig.\ref{Fig3}. The tunneling conductance curves always reproducibly show a fully opened gap of $\Delta_S$=0.28 meV, although the width of the quasiparticle tails slightly changes as a function of the position, indicating local variations of the gap distribution of some tens of $\mu$eV around the value given above (60 $\mu$eV).

\begin{figure}[ht]
\includegraphics[width=8cm,clip]{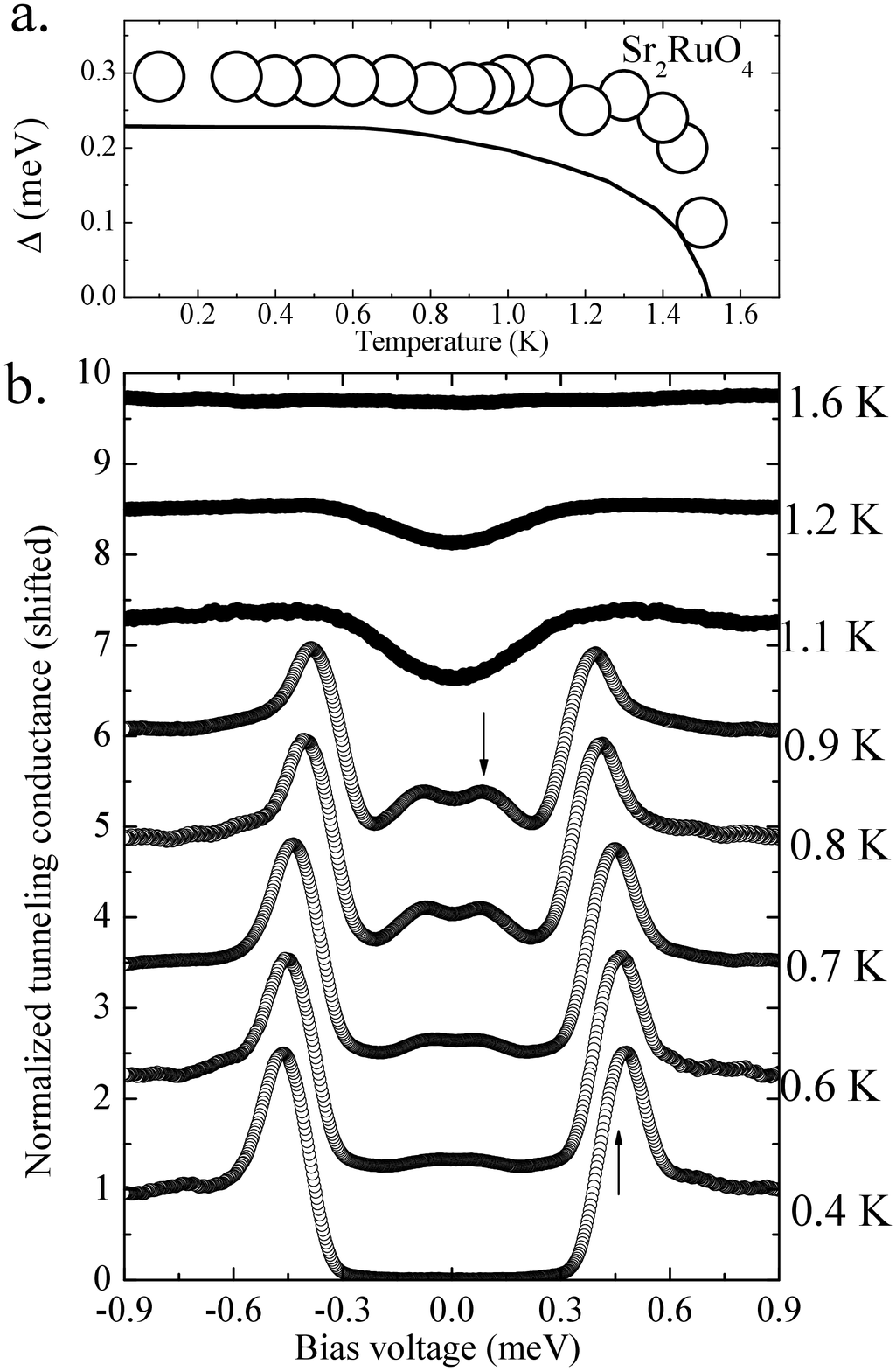}
\vskip -0cm \caption{In a we show the temperature dependence of the superconducting gap of Sr$_2$RuO$_4$ extracted from the temperature dependence of our tunneling conductance curves. Line is the expected temperature dependence from weak coupling s-wave BCS theory. In b we show some representative spectra above 0.4 K. Arrows mark the peaks at $\Delta_S+\Delta_T$ and at $\Delta_S-\Delta_T$.}
\label{Fig4}
\end{figure}

When we increase temperature, we observe the characteristic peaks at $(\Delta_S-\Delta_T)$ (Fig.\ref{Fig4}b) above about 0.5 K. Above 1.1 K, the tip goes into the normal state, and curves corresponding to a normal-superconductor junction are found up to the critical temperature of Sr$_2$RuO$_4$ ($T_c=1.5K$). Below the $T_c$ of Al, we can easily trace $\Delta_S(T)$ of Sr$_2$RuO$_4$ (Fig.\ref{Fig4}a) by just using the information from the position of the peaks (in $\Delta_S-\Delta_T$ and in $\Delta_S+\Delta_T$) in the tunneling conductance as a function of temperature. Above the $T_c$ of Al, we have to use the conventional procedure of unfolding the density of states of the sample from the tunneling conductance\cite{Crespo06a}, and this produces a larger dispersion in $\Delta_S(T)$ of Sr$_2$RuO$_4$. In Fig.\ref{Fig4}a we compare $\Delta_S(T)$ of Sr$_2$RuO$_4$ with the temperature dependence of the gap within BCS theory (line, using $\Delta_0(T=0K)=1.76k_B T_c=$0.229 meV, and $T_c=$1.5 K). The gap magnitude remains about 20\% above predictions of weak couling BCS theory over the whole temperature range. Note that the distribution of values of the superconducting gap discussed above is about the same order. However, $\Delta_S(T)$ shown in Fig.\ref{Fig4} is the center of the distribution, and it is clearly shifted with respect to expectations from BCS theory.

It is out of the scope of the present letter to discuss all available data about the superconducting gap of Sr$_2$RuO$_4$ into agreement. We should however note that several calculations have shown a nearly isotropic gap structure within a triplet pairing scenario (see e.g.\cite{Eschrig01}). Our experiment is indeed fully compatible with the scenario coming out from a number of spin sensitive experiments, which point towards the state described in the $\mathbf{d}$-vector notation with $\mathbf{d}=\Delta_0\mathbf{\hat{z}}(k_x \pm ik_y)$ and leads to an isotropic energy gap\cite{Mackenzie03}. Obviously, our data are also compatible with isotropic s-wave theory with somewhat stronger coupling than BCS theory (to account for the slightly increased $\Delta_S(T)$ of Fig.\ref{Fig4}a). Nevertheless, the amount of experimental data pointing towards an unconventional odd-parity superconductivity scenario seems overwhelming, although it is still not fully definitely  established, as discussed in Ref.\cite{Mackenzie03}.

On the other hand, macroscopic measurements show that the Fermi level superconducting density of states must be zero in pure Sr$_2$RuO$_4$\cite{Nishizaki99,Izawa01,Tanatar01b,Ishida00,Lupien01}, but with a very wide distribution of values of the superconducting gap, compatible with nodes or at least very shallow gap minima at some part of the Fermi surface. At first sight, it may seem incompatible with the present STM observations which point toward a large fully open gap. However, recent STM studies on multigap superconductors have revealed that features of the local tunneling conductance are generally representative of the values of the superconducting gap over only part of the Fermi surface. Possibly, the most prominent example is the case of MgB$_2$, where the features of the superconducting density of states that are most easily resolved in STM are those of the $\pi$ band. The $\sigma$ band features produce a much weaker signal in the tunneling conductance that is difficult to resolve \cite{PhysicaMgB2}. This is easily understood if one considers that the most general expression for the tunneling current must take into account the way tip and sample wave functions couple at the local level\cite{Fischer07}. At present, a detailed association of different gap features observed in STM to different parts of the Fermi surface is only possible through electronic surface wave-function interference effects with defects or impurities\cite{Hoffman02,Fischer07}, and another method requiring detailed atomic resolution Fourier analysis of the tunneling conductance curves is being developed\cite{Guillamon08}. Moreover, in Sr$_2$RuO$_4$, the strong differences in mass renormalization, as well as the weak coupling between different parts of the Fermi surface can easily lead to different superconducting properties, both in gap magnitude and presence of nodes, over different parts of the Fermi surface\cite{Agterberg97,Zhitomirsky01}. In particular, Zhitomirsky and Rice \cite{Zhitomirsky01} discuss the possibility that an unconventional order parameter $\mathbf{d}=\Delta_0\mathbf{\hat{z}}(k_x \pm ik_y)$ with an isotropic gap is found in the Fermi surface band where superconducting correlations are strongest. They find that line nodes can then be induced in the rest of the Fermi surface through a strongly anisotropic proximity effect, intimately connected with the unconventional properties of superconductivity.

In summary, we have found that spin triplet superconductivity in Sr$_2$RuO$_4$ must be associated to a fully opened gap with a value close to expectations from weak coupling BCS theory. This result provides a turning point in the gap structure studies of Sr$_2$RuO$_4$ and calls for more detailed work in this spin triplet superconductor, which clearly has very peculiar gap opening over the Fermi surface.

We acknowledge discussions with A.I. Buzdin, F. Guinea and J.G. Rodrigo. The Laboratorio de Bajas Temperaturas is associated to the ICMM of the CSIC. This work was supported by the Spanish MEC (Consolider Ingenio 2010 and FIS programs), by the Comunidad de Madrid through program "Science and Technology in the Milikelvin", and by NES and ECOM programs of the ESF.


\end{document}